\documentclass[twocolumn,showpacs,preprintnumbers,amsmath,amssymb]{revtex4}

\usepackage{graphicx}
\usepackage{dcolumn}
\usepackage{bm}

\begin{document}


\title{Quantum Transport with Spin Dephasing: 
\\ A Nonequlibrium Green's Function Approach}

\author{Ahmet Ali Yanik}
\email{yanik@purdue.edu}
\affiliation{
Department of Physics,}
\affiliation{
Network for Computational Nanotechnology,\\ Purdue University,
West Lafayette, IN, 47907, USA
}

\author{Gerhard  Klimeck}%
\affiliation{School of Electrical and Computer Engineering,}
\affiliation{Network for Computational Nanotechnology,\\
Purdue University, West Lafayette, IN, 47907, USA
}%
\affiliation{
Jet Propulsion Lab, Caltech, Pasadena, CA, 91109, USA
}%
\author{Supriyo Datta}
\affiliation{
School of Electrical and Computer Engineering}
\affiliation{Network for Computational Nanotechnology,\\
Purdue University, West Lafayette, IN, 47907, USA
}%
\date{\today}

\begin{abstract}
A quantum transport model incorporating  spin scattering processes is presented using 
the non-equilibrium Green's function (NEGF) formalism within the self-consistent Born approximation. 
This model offers a unified approach by capturing the spin-flip scattering and the quantum effects simultaneously. 
A numerical implementation of the model is illustrated for magnetic tunnel junction devices with embedded magnetic
impurity layers. The results are compared with experimental data, revealing the underlying physics of the coherent and 
incoherent transport regimes. It is shown that small variations in magnetic impurity spin-states/concentrations could cause 
large deviations in junction magnetoresistances.
\end{abstract}


\maketitle

\section{\label{sec:level1}INTRODUCTION}

Quantum transport in spintronic devices is currently a topic of great interest. Most of the theoretical work reported so far 
has been based on the Landauer approach \cite{Landauer} assuming coherent transport, although a few authors have included 
incoherent processes through averaging over a large ensemble of disordered configurations \cite{LB-scat1,LB-scat2,LB-scat3}. 
However, it is not straightforward to include dissipative interactions in such approaches. The non-equlibrium Green's function 
(NEGF) formalism provides a natural framework for describing quantum transport in the presence of incoherent and 
dissipative processes. Here, a numerical implementation of the NEGF formalism with spin-flip scattering is presented. For magnetic tunnel 
junctions (MTJs) with embedded magnetic impurity layers, this model is able to capture and explain three 
distinctive experimental features reported in the literature \cite{Moodera1,Moodera2,Davis,Davis1} 
regarding the dependence of the junction magnetoresistances $(JMRs)$ on $(1)$ barrier thickness, $(2)$ barrier height and 
$(3)$ the number of magnetic impurities. The model is quite general and can be used to analyse and design a variety of spintronic 
devices beyond the 1-D geometry explored in this article. 
 
This article is organized as follows. In the view of pedagogical clarity, a heurisic presentation of the NEGF formalism with spin dephasing 
mechanisms is given Sec~\ref{sec:model} followed by a numerical implementation of the model in Sec~\ref{sec:application}. 
Initially (Sec~\ref{sec:coherent}) the definitions of device characteristics are presented for \textit{impurity free} MTJs together with device 
parameters benchmarked against experimental measurements. How to incorporate the spin exchange scattering mechanisms into the electron transport 
calculations is shown (Sec~\ref{sec:spin_scattering}), and the model is applied to MTJ devices with magnetic impurity layers (Sec~\ref{sec:results}). 
Theoretical estimates and experimental measurements are compared as well in this section (Sec~\ref{sec:results}), while a summary of the results is 
given in Sec~\ref{sec:summary}.

\section{\label{sec:model}MODEL DESCRIPTION}

\textit{NEGF Method:} The problem is partitioned into channel and contact regions as illustrated in Fig.~\ref{fig:fig_1} \cite{Datta}. Components of 
the partitioned device can be classified in four categories:

\begin{figure} [t]
\includegraphics [width=85mm,height=40mm]{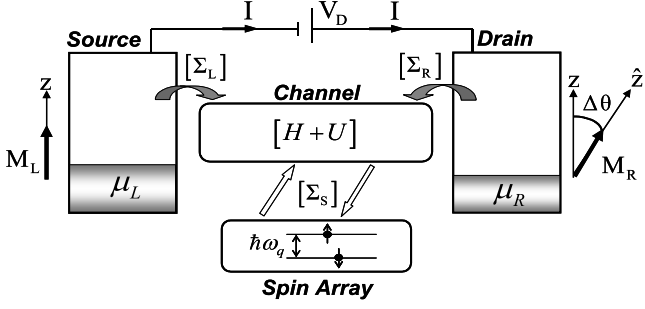}
\caption{\label{fig:fig_1} A schematic illustration of device partitioning in NEGF formalism. 
Magnetization direction of the drain is defined relative to the source(${\rm \Delta \theta  = \theta }_{\rm R} {\rm  - \theta }_{\rm L} $).}
\end{figure}

~(i) \textit{Channel} properties are defined by the Hamiltonian matrix $\left[ H \right]$ including the applied bias potential.

(ii) \textit{Contacts} are included through self-energy matrices ${{\left[ {\Sigma _L } \right]} \mathord{\left/
 {\vphantom {{\left[ {\Sigma _L } \right]} {\left[ {\Sigma _R } \right]}}} \right.
 \kern-\nulldelimiterspace} {\left[ {\Sigma _R } \right]}}$ whose anti-hermitian component:
\begin{equation}
\Gamma _{L,R} \left( E \right) = i\left( {\Sigma _{L,R} \left( E \right) 
- \Sigma _{L,R}^ \dag  \left( E \right)} \right)
,\label{eq:cont_gam}
\end{equation}
describes the broadening due to the coupling to the contact. The corresponding inscattering/outscattering matrices are 
defined as:
\begin{subequations}
\label{eq:cont_scat}
\begin{equation}
\Sigma _{L,R}^{in} \left( E \right) = f_0 \left( {E - \mu _{L,R} } \right)
\Gamma _{L,R} \left( E \right)
,\label{subeq:inscatt}
\end{equation}
\begin{equation}
\Sigma _{L,R}^{out} \left( E \right) = \left[ {1 - f_0 \left( {E - \mu _{L,R} } \right)} \right]\Gamma _{L,R} \left( E \right)
,\label{subeq:outscatt}
\end{equation}
\end{subequations}
where $f_0 \left( {E - \mu _{L,R} } \right) = {1 \mathord{\left/
 {\vphantom {1 {1 + \exp \left[ {{{\left( {E - \mu _{L,R} } \right)} \mathord{\left/
 {\vphantom {{\left( {E - \mu _{L,R} } \right)} {k_B T}}} \right.
 \kern-\nulldelimiterspace} {k_B T}}} \right]}}} \right.
 \kern-\nulldelimiterspace} {1 + \exp \left[ {{{\left( {E - \mu _{L,R} } \right)} \mathord{\left/
 {\vphantom {{\left( {E - \mu _{L,R} } \right)} {k_B T}}} \right.
 \kern-\nulldelimiterspace} {k_B T}}} \right]}}$ is the Fermi function for the related contact.

(iii) \textit{Electron-electron interactions} are incorporated through the mean field electrostatic potential matrix $\left[ U \right]$. 

(iv) \textit{Incoherent} scattering processes in the channel region are described by in/out-scattering matrices 
${{\left[ {\Sigma _S^{in} } \right]} \mathord{\left/
 {\vphantom {{\left[ {\Sigma _S^{in} } \right]} {\left[ {\Sigma _S^{out} } \right]}}} \right.
 \kern-\nulldelimiterspace} {\left[ {\Sigma _S^{out} } \right]}}$. Broadening due to scattering is given by:
\begin{equation}
\Gamma _{\rm S} \left( E \right) = \left[ {\Sigma _S^{in} \left( E \right) + \Sigma _S^{out} \left( E \right)} \right]
,\label{eq:scat_gam}
\end{equation}
from which the self-energy matrix is obtained through a Hilbert transform: 
\begin{equation}
\Sigma _{\rm S} \left( E \right) = \overbrace {\frac{1}{{2\pi }}\int {\frac{{\Gamma _S \left( {E'} \right)}}{{E' - E}}dE'} }^{{\mathop{\rm Re}\nolimits} } - \overbrace {i\frac{{\Gamma _S \left( E \right)}}{2}}^{{\mathop{\rm Im}\nolimits} }
.\label{eq:hilbert}
\end{equation}

Eqs.~(\ref{eq:cont_gam}-\ref{eq:cont_scat}) are the boundary conditions that drive the coupled NEGF equations [Eqs.~(\ref{eq:scat_gam}-\ref{eq:scat_tensor})], where Green's function is defined as:
\begin{equation}
G = \left[ {E I - H - U - \Sigma _L  - \Sigma _R  - \Sigma _S } \right]^{ - 1} 
,\label{eq:green}
\end{equation}
with the spectral function (analogous to density states):
\begin{equation}
A = i\left[ {G - G^ \dag  } \right] = G^n  + G^p 
,\label{eq:spect}
\end{equation}
where ${{\left[ {G^n } \right]} \mathord{\left/ {\vphantom {{\left[ {G^n } \right]} {\left[ {G^p } \right]}}} \right.
 \kern-\nulldelimiterspace} {\left[ {G^p } \right]}}$ refer the \textit{electron/hole correlation functions} (whose diagonal 
elements are the electron/hole density):  
\begin{equation}
G^{n,p}  = G\left[ {\Sigma _L^{in,out}  + \Sigma _R^{in,out}  + \Sigma _S^{in,out} } \right]G^ \dag 
.\label{eq:corr}
\end{equation}

The in/out-scattering matrices ${{\left[ {\Sigma _S^{in} } \right]} \mathord{\left/ {\vphantom {{\left[ {\Sigma _S^{in} } \right]} {\left[ {\Sigma _S^{out} } \right]}}} \right.
 \kern-\nulldelimiterspace} {\left[ {\Sigma _S^{out} } \right]}}$ are related to the electron/hole correlation functions
${{\left[ {G^n } \right]} \mathord{\left/ {\vphantom {{\left[ {G^n } \right]} {\left[ {G^p } \right]}}} \right.
 \kern-\nulldelimiterspace} {\left[ {G^p } \right]}}$ through:
\begin{widetext}
\begin{eqnarray}
\Sigma _{S;\sigma _i \sigma _j }^{in,out} \left( {r,r';E} \right) = \int {\sum\limits_{\sigma _k ,\sigma _l } {\left[ {D_{\sigma _i \sigma _j ;\sigma _k \sigma _l }^{n,p} \left( {r,r';\hbar \omega } \right)} \right]_{sf} G_{\sigma _k \sigma _l }^{n,p} \left( {r,r';E \mp \hbar \omega } \right)d\left( {\hbar \omega } \right)} } \nonumber\\
+ \int {\sum\limits_{\sigma _k ,\sigma _l } {\left[ {D_{\sigma _i \sigma _j ;\sigma _k \sigma _l }^{n,p} \left( {r,r';\hbar \omega } \right)} \right]_{nsf} G_{\sigma _k \sigma _l }^{n,p} \left( {r,r';E} \right)d\left( {\hbar \omega } \right)} }  
.\label{eq:scat_tensor}
\end{eqnarray}
\end{widetext}
Here the spin indices $(\sigma _k ,\sigma _l)$ refer to the (2x2) block diagonal elements of the on-site electron/hole correlation 
function which is related through the $[D^n]/[D^p]$ tensors to the $(\sigma _i ,\sigma _j)$ spin components of the (2x2) block diagonal of the in/out-scattering function. 
This term can be interpreted as in the following. The first part describes the process of 
spin-flip transitions (subscript sf) due to the spin-exchange scatterings in the channel region. The second part denotes the contributions of the spin-conserving exchange 
scatterings (subscript nsf for "no spin-flip") in the channel region. Both of the contributing parts are previously shown by Appelbaum\cite{Appelbaum} using a similar treatment. Here the $[D^n]/[D^p]$ are fourth-order 
scattering tensors, describing the spatial correlation and the energy spectrum of the underlying microscopic \textit{spin-dephasing} scattering mechanisms. 
These scattering tensors can be obtained from the spin scattering hamiltonian:

\begin{equation}
H_{{\mathop{\rm I}} } \left( {\vec r} \right) = \sum\nolimits_{R_{j} } {J\left( {\vec r - \vec R_{j} } \right)\vec \sigma  \cdot \vec S_{j} } 
,\label{eq:hamiltonian}
\end{equation}
where ${\vec r}$/${\vec R_{j} }$ are the spatial coordinates and ${\vec \sigma }$/${\vec S_{j} }$ are the spin operators for the \textit{channel electron/(j-th) 
magnetic-impurity}. For point like exchange scattering processes, the scattering tensor for the first term 
in Eq.~(\ref{eq:scat_tensor}) corresponding to the spin-flip transitions is given by (see Appendix):

\begin{widetext}
\begin{equation}
\begin{array}{*{20}c}
   \begin{array}{l}
 \left| {\sigma _k \sigma _l } \right\rangle  \to  \\ 
 \text{~~~~~} \left\langle {\sigma _i \sigma _j } \right| \downarrow  \\ 
 \end{array} & {\begin{array}{*{20}c}
\text{~~~~~~~~}
   {\left| { \uparrow  \uparrow } \right\rangle }               \text{~~~~~~~~~~~~~~}
& {\left| { \downarrow  \downarrow } \right\rangle }            \text{~~~~~}
& {\left| { \uparrow  \downarrow } \right\rangle }              \text{}
& {\left| { \downarrow  \uparrow } \right\rangle }  \\          \text{}
\end{array}}  \\
   {\left[ {{\rm D}^{{\rm n,p}} \left( {r,r';\hbar \omega } \right)} \right]_{{\rm sf}} {\rm   } = \delta(r - r') \sum\limits_{\omega _q } {J^2 N{}_I(\omega_q)} {\rm } \text{~~} \begin{array}{*{20}c}
   {\left\langle { \uparrow  \uparrow } \right| }  \\
   {\left\langle { \downarrow  \downarrow } \right| }  \\
   {\left\langle { \uparrow  \downarrow } \right| }  \\
   {\left\langle { \downarrow  \uparrow } \right| }  \\
\end{array}} & {\left[ {\begin{array}{*{20}c}
   0 & {F_{u,d} \delta \left( {\omega  \mp \omega _q } \right)} & 0 \text{~~~~} & 0  \\
   {F_{d,u} \delta \left( {\omega  \pm \omega _q } \right)} & 0 & 0 \text{~~~~} & 0  \\
   0 & 0 & 0 \text{~~~~}& 0  \\
   0 & 0 & 0 \text{~~~~} & 0  \\
\end{array}} \right]}  \\
\end{array}
.\label{eq:spin_flip}
\end{equation}
\end{widetext}
Also, the no-spin flip component in Eq.~(\ref{eq:scat_tensor}) is given by:

\begin{widetext}
\begin{equation}
\begin{array}{*{20}c}
      \begin{array}{l}
 \left| {\sigma _k \sigma _l } \right\rangle  \to  \\ 
 \text{~~~~~} \left\langle {\sigma _i \sigma _j } \right| \downarrow  \\ 
 \end{array} & {\begin{array}{*{20}r}
 \text{~~~}  
   {\left| { \uparrow  \uparrow } \right\rangle }             \text{~}  
& {\left| { \downarrow  \downarrow } \right\rangle }          \text{~}  
& {\left| { \uparrow  \downarrow } \right\rangle }            \text{~}  
& {\left| { \downarrow  \uparrow } \right\rangle }  \\        \text{~}  
\end{array}}  \\
   {\left[ {{\rm D}^{{\rm n,p}} \left( {r,r';\hbar \omega } \right)} \right]_{{\rm nsf}} {\rm   } = \delta(r - r') \sum\limits_{\omega_q} {\frac{1}{4}} J^2 N{}_I(\omega_q) [\delta \left( {\omega  - \omega _q } \right)] {\rm  } \text{~~} \begin{array}{*{20}c}
{\left\langle { \uparrow  \uparrow } \right| }  \\
{\left\langle { \downarrow  \downarrow } \right| }  \\
{\left\langle { \uparrow  \downarrow } \right| }  \\
{\left\langle { \downarrow  \uparrow } \right| }  \\
\end{array}} & {\left[ {\begin{array}{*{20}r}
   \text{~~~} {\rm 1} & \text{~~~}0 & \text{~~~}0 & \text{~~~}0  \\
   \text{~~~}0 & \text{~~~}1 & \text{~~~}0 & \text{~~~}0  \\
   \text{~~~}0 & \text{~~~}0 & \text{~~~}{ - 1} & \text{~~~}0  \\
   \text{~~~}0 & \text{~~~}0 & \text{~~~}0 & \text{~~~}{ - 1}  \\
\end{array}} \right]}  \\
\end{array} 
.\label{eq:no_spin_scat}
\end{equation}
\end{widetext}

\textit{Current} is calculated from the self-consistent solution of the above equations for any terminal \textit{"i"}:
\begin{equation}
I_i  = \frac{q}{h}\int\limits_{ - \infty }^\infty  {trace\left( {\left[ {\Sigma _i^{in} \left( E \right)A\left( E \right)} \right] - \left[ {\Gamma _i \left( E \right)G^n \left( E \right)} \right]} \right)} {\rm  }dE ) 
.\label{eq:curr}
\end{equation}

The general solution scheme without going into the details can be summarized as follows . The matrices listed under the categories (i) and (ii) are fixed at the outset 
of any calculations. While the $[U]$, $[\Sigma _S^{in,out} ]$ and $\left[ {\Sigma _S } \right]$ matrices under the 
\textit{charging and scattering} categories (iii) and (iv) depend on the correlation and spectral 
functions requiring an \textit{iterative self-consisted solution} of the NEGF Equations [Eqs.~(\ref{eq:scat_gam}-\ref{eq:scat_tensor})]. 
One important thing to note is that for the numerical implementation presented in Sec~\ref{sec:application} we do not compute the charging potential $[U]$ 
self-consistently with the charge. The change in tunnel barriers is neglected and assumed not to influence the electrostatic potential. 
This allows one to focus on the \textit{dephasing} due to the spin-flip interactions.

\section{\label{sec:application}APPLICATION: MTJ\lowercase{s} WITH MAGNETIC IMPURITIES}

In the presence of "rigid" scatterers such as impurities and defects, electron transport is considered 
coherent as the phase relationships between different paths are time independent. Accordingly, any static 
deviations from perfect crystallinity leading to the scattering of the electrons from one state to another 
can be incorporated into the transport problem through the device Hamiltonian H. 
However the situtation is different when the impurities have an internal degree of freedom,  
such as fluctuating internal spin states in the case of magnetic impurities. Electron scatterings from such 
impurities randomize the electron spin states which can not be simply incorporated through the device 
Hamiltonian. Instead scattering self energy matrices are needed for this type of phase-relaxing scattering proccesses. 
An implementation of this self-energy matrix treatment will be discussed in this 
part of the paper for electron-impurity exchange scattering processes in MTJs. But first, we'll discuss 
MTJ fundamentals and device characteristics in the absence of magnetic impurities \textit{(the coherent regime)}.

\begin{figure} [b]
\includegraphics [width=85mm,height=33mm]{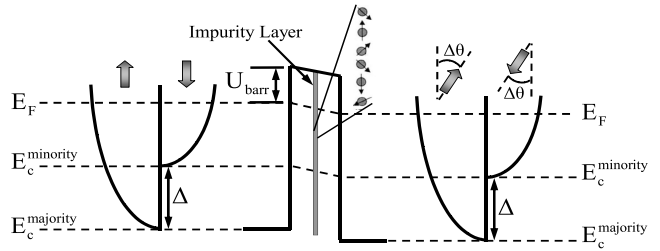}
\caption{\label{fig:fig_2} Energy band diagram for the model MTJs considered here.}
\end{figure}

\subsection{\label{sec:coherent} MTJs: Coherent Regime}

MTJs devices considered here consist of a tunneling barrier (${\rm AlO}_{\rm x}$) sandwiched between two ferromagnets (Co) with different 
magnetic coercivities enabling independent manipulation of contact magnetization directions (Fig.~\ref{fig:fig_2}). Single band 
tight-binding approximation is adopted \cite{Stearns} with an effective electron mass ($m^*=m_e$) in the tunneling region and the ferromagnetic contacts. 
Accordingly for constant effective mass throughout the device, transverse modes can be included using 2-D integrated Fermi functions $f_{2D} \left( {E_z  - \mu _{L,R} } \right) = \left( {{{m^* k_B T} \mathord{\left/
 {\vphantom {{m^* k_B T} {2\pi \rlap{--} h^2 }}} \right. \kern-\nulldelimiterspace} {2\pi \rlap{--} h^2 }}} \right)\ln \left[ {1 + \exp \left( {{{\mu _{L,R}  - E_z } \mathord{\left/
 {\vphantom {{\mu _{L,R}  - E_z } {k_B T}}} \right. \kern-\nulldelimiterspace} {k_B T}}} \right)} \right]$ in Eq.~(\ref{eq:cont_scat}) instead of numerically summing parallel k-components. 
The Green's function of the device is simply defined as:
\begin{equation}
G(E_z)= \left( {E_zI - H - \Sigma_L - \Sigma_R } \right)^{ - 1} 
,\label{eq:green_coherent}
\end{equation}
without any self-consistent solution requirements where H is Hamiltonian of the isolated system, and $\Sigma_L/\Sigma_R $ are the self-energies due to the source/drain contacts. 
Here the matrices have twice the size of the channel region in the corresponding presentation due to the electron spin states. 
Accordingly, in real space representation for a discrete lattice whose points are located at $x = ja$, j being an integer ($j = 1 \cdots N$), the matrix $({E_zI - H - \Sigma_L - \Sigma_R })$ can be expressed as:
\begin{widetext}
\begin{equation}
\begin{array}{*{20}c}
   {}  \\
   {}  \\
   {}  \\
   {E_zI - H - \Sigma_L - \Sigma_R  = }  \\
   {}  \\
   {}  \\
\end{array}\begin{array}{*{20}c}
\text{}
   {} & {\begin{array}{*{20}r}                    
   {{\rm  }\left| 1 \right\rangle }               \text{~~~~~~~~~~~}
& {{\rm   }\left| 2 \right\rangle }               \text{~~~~}
& { }                                             \text{~~~~~~}
& {{\rm  }\left| {N - 1} \right\rangle }          \text{~~~~~~~~~~}
& {\left| N \right\rangle }  \\                   \text{~~~~~~~~~~~~~~}
\end{array}}  \\
   {\begin{array}{*{20}c}
   {\left\langle 1 \right| }  \\
   {\left\langle 2 \right| }  \\
   {} \\
   {\left\langle {N-1} \right| }  \\
   {\left\langle {N} \right| }  \\
\end{array}} & {\left[ {\begin{array}{*{20}c}
   {E_z \bar I - \alpha _1  - \bar \Sigma _L } & \beta  &  \cdots  & {\bar 0} & {\bar 0}  \\
   {\beta ^ +  } & {E_z \bar I - \alpha _2 } &  \cdots  & {\bar 0} & {\bar 0}  \\
    \vdots  &  \vdots  &  \ddots  &  \vdots  &  \vdots   \\
   {\bar 0} & {\bar 0} &  \cdots  & {E_z \bar I - \alpha _{N - 1} } & \beta   \\
   {\bar 0} & {\bar 0} &  \cdots  & {\beta ^ +  } & {E_z \bar I - \alpha _N  - \bar \Sigma _R }  \\
\end{array}} \right]}  \\
\end{array}
,\label{eq:tri_diag}
\end{equation}
\end{widetext}
where ${\alpha_n}$ is a 2x2 on-site matrix:
\begin{equation}
\alpha _n  = \left[ {\begin{array}{*{20}c}
   {E_{c,n}^ \uparrow   + 2t + U_n } & 0  \\
   0 & {E_{c,n}^ \downarrow   + 2t + U_n }  \\
\end{array}} \right]
,\label{eq:alfa}
\end{equation}
and $\beta  =  - t \bar I$ is a 2x2 site-coupling matrix with $t = {{\hbar ^2 } \mathord{\left/ {\vphantom {{\hbar ^2 } {2m}}} \right. \kern-\nulldelimiterspace} {2m}}a^2 $ and ${\bar I} = ( {\begin{array}{*{20}c}1 & 0  \\ 0 & 1  \\ \end{array}} )$. 
The left contact self-energy matrix is nonzero only for the first 2x2 block: 

\begin{equation}
\Sigma _L \left( {1,1;E_z } \right) = \bar \Sigma _L = \left[ {\begin{array}{*{20}c}
   { - te^{ik_L^ \uparrow  a} } & 0  \\
   0 & { - te^{ik_L^ \downarrow  a} }  \\
\end{array}} \right]
,\label{eq:self_left}
\end{equation}
where $E_z  = E_c^{ \uparrow , \downarrow }  + U_{L}  + 2t\left( {1 - \cos {\rm  }k_{L}^{ \uparrow , \downarrow } a} \right)$. 
For the right contact only the last block is non-zero:
\begin{equation}
\Sigma _R \left( {N,N,E_z } \right) = \bar \Sigma _R = \tilde \Re \left[ {\begin{array}{*{20}c}
   { - te^{ik_R^ \uparrow  a} } & 0  \\
   0 & { - te^{ik_R^ \downarrow  a} }  \\
\end{array}} \right]\tilde \Re^\dag 
,\label{eq:unitary}
\end{equation}
where $E_z  = E_c^{ \uparrow , \downarrow }  + U_{R}  + 2t\left( {1 - \cos {\rm  }k_{R}^{ \uparrow , \downarrow } a} \right)$ with ${\tilde\Re}$ being 
the unitary rotation operator defined as:
\begin{equation} 
\tilde \Re\left( {\Delta \theta } \right) = \left[ {\begin{array}{*{20}r}
   {\cos \left( {{{\Delta \theta } \mathord{\left/
 {\vphantom {{\Delta \theta } 2}} \right.
 \kern-\nulldelimiterspace} 2}} \right)} & {\sin \left( {{{\Delta \theta } \mathord{\left/
 {\vphantom {{\Delta \theta } 2}} \right.
 \kern-\nulldelimiterspace} 2}} \right)}  \\
   { - \sin \left( {{{\Delta \theta } \mathord{\left/
 {\vphantom {{\Delta \theta } 2}} \right.
 \kern-\nulldelimiterspace} 2}} \right)} & {\cos \left( {{{\Delta \theta } \mathord{\left/
 {\vphantom {{\Delta \theta } 2}} \right.
 \kern-\nulldelimiterspace} 2}} \right)}  \\
\end{array}} \right]
,\label{eq:unitary}
\end{equation}
for two contacts with magnetization directions differing by an angle $\Delta \theta$.

\textit{A theoretical analysis} of MTJ devices in the absence of magnetic impurity layers is presented and compared with the 
experimental data \cite{Moodera1,Davis} for varying tunneling barrier heights and thicknesses.   
The parameters used here for the generic ferromagnetic contacts are the Fermi energy $E_F  = 2.2$ eV and the exchange field 
$\Delta  = 1.45$ eV \cite{Stearns}. The tunneling region potential barrier $\left[ {U_{barr} } \right]$ is parameterized 
within the band gaps quoted from the literature \cite{Moodera_Initial,Henrich}, while the charging potential $\left[ {U} \right]$ is neglected 
due to the pure tunneling nature of the transport.

\textit{Coherent} tunneling regime features are obtained by benchmarking the experimental measurements made in impurity 
free tunneling oxide MTJs at small bias voltages.  Referring to ${{I_F } \mathord{\left/{\vphantom {{I_F } {I_{AF} }}} \right. \kern-\nulldelimiterspace} {I_{AF} }}$ 
as the current values for the parallel/antiparallel magnetizations (${{\Delta \theta  = 0} \mathord{\left/
{\vphantom {{\Delta \theta  = 0} {\Delta \theta  = \pi }}} \right. \kern-\nulldelimiterspace} {\Delta \theta  = \pi }}$)
of the ferromagnetic contacts, the $JMR$ is defined as:   

\begin{equation}
JMR = {{\left( {I_F  - I_{AF} } \right)} \mathord{\left/
 {\vphantom {{\left( {I_F  - I_{AF} } \right)} {I_F }}} \right.
 \kern-\nulldelimiterspace} {I_F }}
.\label{eq:JMR}
\end{equation}

\begin{figure} [t]
\includegraphics [width=85mm,height=60mm]{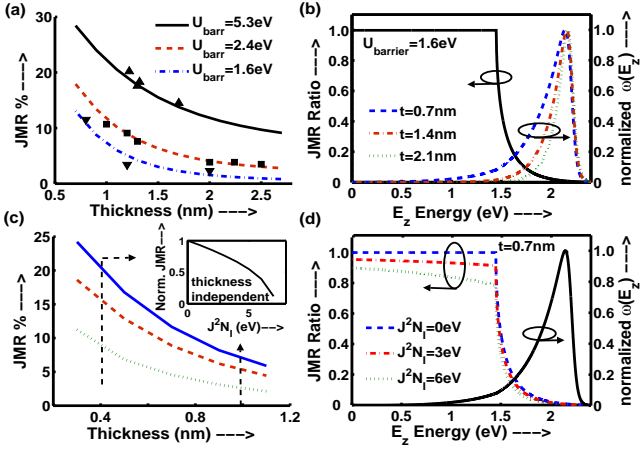}
\caption{\label{fig:fig_3} For impurity free MTJs, (a)thickness dependence of $JMRs$ for different barrier heights are shown in comparison 
with experimental measurements \cite{Matsuda,Chen,Zhang,Mitsuzuka,Oepts,Beech,Yamanaka,Moodera3,Sun,Shang} while an (b)energy resolved analysis of
$JMR\left( {E_z } \right)$ \textit{(left-axis)} and \textit{normalized} $w\left( {E_z } \right)$ (\textit{right-axis}) distributions is also presented 
for a device with a tunneling barrier height of \textit{1.6 eV}. For MTJs with impurity layers, (c)variation of $JMRs$ 
for varying barrier thicknesses and interactions strengths are shown together with (d)an energy resolved analysis. Normalized 
$JMRs$ are proven to be thickness independent as displayed in the inset.}
\end{figure}

The dependence of the $JMRs$ on the thickness and the height 
of the tunneling barriers is shown in Fig.~\ref{fig:fig_3}(a) with an energy resolved analysis [Fig.~\ref{fig:fig_3}(b)] for 
different barrier thicknesses ($0.7-1.4-2.1$ nm). $JMR$ values are shown to be improving with increasing barrier heights for 
all barrier thicknesses, a  theoretically predicted \cite{Davis,Davis1,Mathon} and experimentally observed \cite{Matsuda,Chen,Zhang,Mitsuzuka,Oepts,Beech,Yamanaka,Moodera3,Sun,Shang} 
feature in MTJs. The barrier heights obtained here may differ from those reported in literature \cite{Matsuda,Chen,Zhang,Mitsuzuka,Oepts,Beech,Yamanaka,Moodera3,Sun,Shang} 
based on empirical models \cite{Simmons}. 

Experiments and theoretical calculations observe deterioration of $JMRs$ with increasing barrier thicknesses [Fig.~\ref{fig:fig_3}(a)]. 
Whereas an energy resolved theoretical analysis shows that energy by energy junction magnetoresistances defined as:
\begin{equation}
JMR(E_z)=({I_F(E_z)-I_{AF}(E_z)}) / I_F(E_z)
,\label{eq:JMR_Ez}
\end{equation}
remain unchanged [Fig.~\ref{fig:fig_3}(b)].  This initially counter intuitive observation can be understood by considering the redistribution of tunneling electron 
densities over energies with changing tunneling barrier thicknesses.  
Defining $w\left( {E_z } \right)$ as a measure of the contributing weight of the $JMR(E_z)$, one can show that experimentally measured $JMR$ is a weighted integral of $JMR(E_z)$s over $E_z$ energies: 

\begin{equation}
JMR = \int {\omega \left( {E_z } \right)JMR\left( {E_z } \right){\rm  }dE_z } 
,\label{eq:JMR_define}
\end{equation}
where $\omega \left( {E_z } \right) = {{I_F \left( {E_z } \right)} \mathord{\left/ {\vphantom {{I_F \left( {E_z } \right)} {I_F }}} \right. \kern-\nulldelimiterspace} {I_F }}$ 
is the energy resolved \textit{spin-continuum current component} (weighting function). In Eq.~(\ref{eq:JMR_define}), independently from the barrier thicknesses  $JMR\left( {E_z } \right)$ ratios are \textit{constant} 
(solid line in Fig.~\ref{fig:fig_3}(b)), while the \textit{normalized} $w\left( {E_z } \right)$ distributions shifts towards higher energies with increasing barrier thicknesses 
(dashed lines in Fig.~\ref{fig:fig_3}(b)). Hence $JMRs$, an integral of the multiplication of the $w\left( {E_z } \right)$ distributions  
with the energy resolved $JMR\left( {E_z } \right)$s, decreases with increasing barrier thicknesses (Eq.~(\ref{eq:JMR})).

\subsection{\label{sec:spin_scattering} Adding Spin Exchange Scattering}

Spin exchange scattering processes are responsible from the incoherent nature of the tunneling transport for the model devices considered here. 
Through this elastic scattering process, the state of nothing else seem to change if the electron and the impurity spins are considered as a composite 
system. Nevertheless, it is an incoherent process since the state of the impurity has changed. What makes this process incoherent are the external forces 
forcing the impurity spins into local equilibrium. The incoherent nature of the scattering lies in the "information erasure" of the 
surrounding through the forces forcing the impurity spins into an unpolarized (\%50 up, \%50 down) spin distribution. These external forces in a 
closely packed impurity layer can be magnetic dipole-dipole interactions among the magnetic impurities or spin relaxation processes coupled with phononic 
excitations. Nevertheless the physical origin of the equilibrium restoring processes is not of our interest (at least from tunneling electron's 
point of view) assuming equilibrium restoring processes are fast enough to maintain the impurity spins in a thermal equilibrium state. Accordingly, NEGF formalism 
incorporates spin dephasing effects of the environment into the electron transport problem through a boundary condition (the spin-scattering self-energy). 
As discussed in Sec~\ref{sec:model}, coupling between the number of available electrons/holes ($ {{\left[ {G^n } \right]} \mathord{\left/ 
{\vphantom {{\left[ {G^n } \right]} {\left[ {G^p } \right]}}} \right. \kern-\nulldelimiterspace} {\left[ {G^p } \right]}} $) at a state and the 
in/out-flow ($ {{\left[ {\Sigma _S^{in} } \right]} \mathord{\left/ {\vphantom {{\left[ {\Sigma _S^{in} } \right]} {\left[ {\Sigma _S^{out} } \right]}}} \right. 
\kern-\nulldelimiterspace} {\left[ {\Sigma _S^{out} } \right]}} $) to/from that state  is related through the fourth order scattering tensor $ {{\left[ {D^n } \right]} \mathord{\left/ 
{\vphantom {{\left[ {D^n } \right]} {\left[ {D^p } \right]}}} \right. \kern-\nulldelimiterspace} {\left[ {D^p } \right]}} $ in Eq.~(\ref{eq:scat_tensor}).

For the model systems considered here magnetic impurity spin states are degenerate ($\hbar \omega_q  = 0$) allowing only elastic spin-flip transitions. 
Spin-conserving scattering processes are also neglected due to their minor effect on the JMRs. Accordingly, the spin scattering tensor relationship given 
in Eqs.~(\ref{eq:scat_tensor},\ref{eq:spin_flip} and \ref{eq:no_spin_scat}) will simplify to:

\begin{equation}
\Sigma _S^{in,out} \left( {r,r;E} \right) = \left[ {D_{\sigma _i \sigma _j ;\sigma _k \sigma _l }^{n,p} } \right]_{sf} G_{\sigma _k \sigma _l }^{n,p} \left( {r,r;E} \right)
.\label{eq:dephasing_tensor}
\end{equation}

${{\left[ {D^n } \right]} \mathord{\left/ {\vphantom {{\left[ {D^n } \right]} {\left[ {D^p } \right]}}} \right. \kern-\nulldelimiterspace} {\left[ {D^p } \right]}}$ 
scattering tensors relate the electron/hole $ {{\left[ {G^n } \right]} \mathord{\left/ 
{\vphantom {{\left[ {G^n } \right]} {\left[ {G^p } \right]}}} \right. \kern-\nulldelimiterspace} {\left[ {G^p } \right]}} $ correlation matrices with the 
(2x2) block diagonal elements of $ {{\left[ {\Sigma _S^{in} } \right]} \mathord{\left/ {\vphantom {{\left[ {\Sigma _S^{in} } \right]} {\left[ {\Sigma _S^{out} } \right]}}} \right. 
\kern-\nulldelimiterspace} {\left[ {\Sigma _S^{out} } \right]}} $ in/out-scattering matrices of form:

\begin{widetext}
\begin{equation}
\Sigma _S^{in,out}  = \left( {\begin{array}{*{20}c}
   {\left( {\bar \Sigma ^{in,out}_S } \right)_{1,1} } & {\bar 0} &  \cdots  & {\bar 0}  \\
   {\bar 0} & {\left( {\bar \Sigma^{in,out}_S } \right)_{2,2} } &  \cdots  & {\bar 0}  \\
    \vdots  &  \vdots  &  \ddots  &  \vdots   \\
   {\bar 0} & {\bar 0} &  \cdots  & {\left( {\bar \Sigma^{in,out}_S } \right)_{N,N} }  \\
\end{array}} \right)
,\label{eq:scattering_function}
\end{equation}
\end{widetext}
through the tensor relationship \textit{(shown below in matrix format)} for the corresponding lattice site "j" with magnetic impurities:

\begin{equation}
\left[ {\begin{array}{*{20}c}
   {\left( {\Sigma _{S; \uparrow  \uparrow }^{in,out} } \right)_{jj} }  \\
   {\left( {\Sigma _{S; \downarrow  \downarrow }^{in,out} } \right)_{jj} }  \\
   {\left( {\Sigma _{S; \uparrow  \downarrow }^{in,out} } \right)_{jj} }  \\
   {\left( {\Sigma _{S; \downarrow  \uparrow }^{in,out} } \right)_{jj} }  \\
\end{array}} \right] = \overbrace {J^2 N_I \left[ {\begin{array}{*{20}c}
   0 & {F_{u,d} } & 0 & 0  \\
   {F_{d,u} } & 0 & 0 & 0  \\
   0 & 0 & 0 & 0  \\
   0 & 0 & 0 & 0  \\
\end{array}} \right]}^{\left[ {D^{n,p} } \right]_{sf}} \left[ {\begin{array}{*{20}c}
   {\left( {G_{ \uparrow  \uparrow }^{n,p} } \right)_{jj} }  \\
   {\left( {G_{ \downarrow  \downarrow }^{n,p} } \right)_{jj} }  \\
   {\left( {G_{ \uparrow  \downarrow }^{n,p} } \right)_{jj} }  \\
   {\left( {G_{ \downarrow  \uparrow }^{n,p} } \right)_{jj} }  \\
\end{array}} \right]
,\label{eq:tensor_relation}
\end{equation}
where $N_I$  is the number of magnetic impurities and $F_u / F_d$ represents fractions of spin-up/spin-down impurities for an uncorrelated ensemble ($F_u  + F_d  = 1$). 

This tensor relationship can be understood heuristically from elementary arguments. The in/out-scattering into 
\textit{spin-up} component is proportional to the density of the \textit{spin-down electrons/holes} times the number of \textit{spin-up impurities}, 
$N_I F_u $ :
\begin{equation}
\left( {\Sigma _{S, \uparrow  \uparrow }^{in,out} } \right)_{jj}  = J^2 N_I F_u \left( {G_{ \downarrow  \downarrow }^{n,p} } \right)_{jj}\\
.\label{eq:ioscat_up}
\end{equation}

Similarly, the in/out-scattering into \textit{spin-down} component is proportional to the density of the \textit{spin-up electrons/holes} times the 
number of 
\textit{spin-down impurities}, $N_I F_d $ :
\begin{equation}
\left( {\Sigma _{S, \downarrow  \downarrow }^{in,out} } \right)_{jj}  = J^2 N_I F_d \left( {G_{ \uparrow  \uparrow }^{n,p} } \right)_{jj}  \\ 
.\label{eq:ioscat_down}
\end{equation}

\subsection{\label{sec:results} Incoherent Regime: Results }

The \textit{incoherent} tunneling regime device characteristics in the presence of magnetic impurities is studied for a fixed barrier height $U_{barr}  = 1.6$ eV [Fig.~\ref{fig:fig_3} (c)] with changing barrier thicknesses 
and electron-impurity spin exchange interactions ($J^2 N_I = 0 - 6$ eV). \textit{Nonlinear} decreasing $JMRs$ with increasing spin-exchange interactions are observed at all barrier thicknesses due to the mixing 
of independent spin-channels \cite{Moodera1,Moodera2} while the \textit{normalized} $JMRs$ are proven to be thickness independent \textit{(inset)}. This observation 
is attributed to the elastic nature of the spin exchange interactions yielding a total drop in $JMR\left( {E_z } \right)$ values at all ${E_z }$ energies in Eq.~(\ref{eq:JMR}) 
while preserving the \textit{normalized} $\omega \left( {E_z } \right)$ carrier distributions [Fig.~\ref{fig:fig_3}(d)]. 

\begin{figure} [b]
\includegraphics [width=85mm,height=60mm]{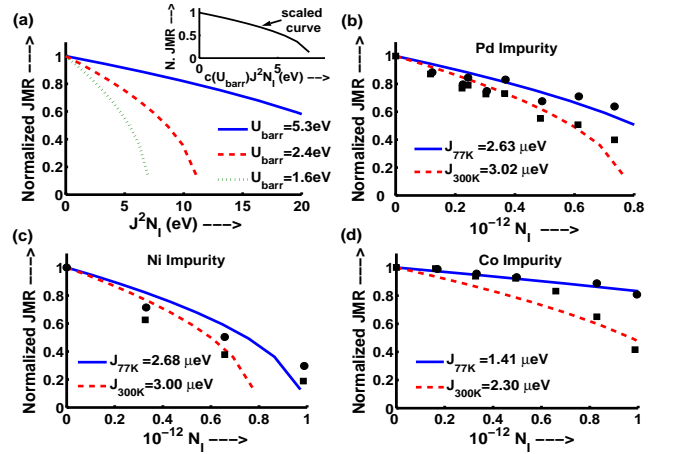}
\caption{\label{fig:fig_4} (a) \textit{Normalized} $JMRs$ deteriorates with increasing spin-dephasing strengths independently from the tunneling barrier 
heights. This general trend can be scaled to a \textit{single universal curve} (inset). Experimental data taken at $77 K$ and $300 K$ is compared with 
theoretical analysis in the presence of (b) Pd, (c) Ni and (d) Co magnetic impurities with increasing impurity concentrations.}
\end{figure}

Further analysis shows that \textit{"normalized"} $JMR$s deteriorates with increasing spin-dephasing strenghts ($J^2N_I$) independently 
from the tunneling barrier heights [Fig.~\ref{fig:fig_4}(a)]. This general trend can be shown by mapping the \textit{"normalized"} $JMR$s into 
a \textit{single universal curve} using a tunneling barrier height dependent scaling constant $c\left( {U_{barr} } \right)$ (inset in Fig.~\ref{fig:fig_4}(a)). 

This allows us to choose a particular value of barrier height [$U_{barr}  = 1.6eV $] and adjust a \textit{single parameter} $J$ to fit our NEGF 
calculations [Fig.~\ref{fig:fig_4}(b-d)] with experimental measurements obtained from ${\rm \delta }$-doped MTJs \cite{Moodera1}.
Submonolayer impurity thicknesses given in the measurements are converted into number of impurities using material concentrations of Pd/Ni/Co 
impurities and device cross sections ($6 \times 10^{ - 4} ~{\rm  cm}^{\rm 2} $) \cite{Moodera1}. 

\textit{Close fitting} to the experimental data are observed at $77 K$ [Fig.~\ref{fig:fig_4}(b-d)] using physically reasonable exchange coupling constants of 
$J=2.63\mu eV/ 2.68\mu eV/ 1.41\mu eV$ for devices with Pd/Ni/Co impurities \cite{Kane}. However, experimentally observed temperature dependence 
of \textit{normalized} $JMR$ ratios can not be accounted for by our model calculations. Broadenings of the electrode Fermi distributions due to 
changing temperatures from $77 K$ to $300 K$ seem to yield variations in \textit{normalized} $JMR$ ratios within a linewidth.  As a result, different 
$J$ exchange couplings are used in order to match the experimental data taken at 300K. 

For \textit{Pd and Ni doped} MTJs, relatively small variations in $J$ exchange couplings are needed (${J_{300}/J_{77}} = 1.16$ for Pd and ${J_{300}/J_{77}} = 1.19$ for Ni) 
in order to match the experimental data at 300K. These small temperature dependences could be due to the presence of some secondary mechanisms not included in our 
calculations. One such mechanism reported in literature includes the presence of impurity-assisted conductance contribution through the defects \textit{(possibly created by the inclusion of 
magnetic impurities within the barrier)} which is known to be strongly temperature dependent \cite{sousa}. In fact, the contribution of the 
impurity-assistant conductance is proportional with the impurity concentrations in accordance with the experimental measurements. 

Another interesting feature observed in calculations [Fig.~\ref{fig:fig_4}(b-c)] is the comparable J exchange couplings for the Pd and Ni impurities 
at temperatures 77K and 300K.  Accordingly, a possible estimate of the impurity spin states may be made by considering the most commonly encountered oxidation 
states of the Pd and Ni impurities. Closed-shell elemental Pd is only known to be in a magnetic oxidized state of S=1 in octahedral oxygen coordination 
according to the Hund's rules \cite{Cox}. Similarly, we may attribute the comparable J couplings for Ni impurities due to the S=1 spin state of the ${\rm Ni}^{+2}$ which 
is known to be a frequently observed ionized state in oxygen environment\cite{Geschwind}.

On the contrary, for \textit{Co doped} MTJs, there's a clear distinction for \textit{normalized} $JMR$ ratios at different temperatures 
[Fig.~\ref{fig:fig_4}(d)] which can not be justified by the presence of secondary mechanisms. 
Fitting these large deviations require large variations in $J$ exchange coupling parameters [${{J_{300} } \mathord{\left/ {\vphantom {{J_{300} } {J_{77} }}} \right. \kern-\nulldelimiterspace} {J_{77} }} = 1.73$].   
We propose this to be a result of thermally driven \textit{low-spin/high-spin phase} transition \cite{phase1,phase2}. This is a credible argument since the oxidation state of the cobalt 
atoms can be in ${\rm Co}^{{\rm  + 2}}$ ($S=3/2$, high-spin) or ${\rm Co}^{{\rm  + 3}}$ ($S=0$, low-spin) state or partially in both of the states 
depending on the oxidation environment. Such thermally driven low-spin/high-spin phase transitions for metal-oxides have been predicted by theoretical calculations and 
observed in experimental studies \cite{phase1,phase2}. These phase transitions have not been discussed in MTJs community in connection with possible scattering 
factors determining the temperature dependence of $JMRs$. Although from the avaliable experimental data it's not possible to make a decisive conclusion in this direction, 
given the \textit{nonlinear} dependence of $JMRs$ on magnetic impurity states in our calculations, we believe that it's important to point out this possibility here.

\section{\label{sec:summary}SUMMARY}
\textit{Summary:} A NEGF based quantum transport model incorporating spin-flip scattering processes within the self-consistent Born approximation is presented. 
Spin-flip scattering and quantum effects are simultaneously captured. Spin scattering operators are derived for the specific case of
electron-impurity spin-exchange interactions and the formalism is applied to spin dependent electron transport in MTJs with magnetic impurity layers. The 
theory is benchmarked against experimental data involving both coherent and incoherent transport regimes. $JMRs$ are shown to decrease both with 
barrier thickness and with spin-flip scattering but our unified treatment clearly brings out the difference in the underlying physics [Fig.~\ref{fig:fig_3}]. 
Our numerical results show that both barrier height and the exchange interaction constant $J$ can be subsumed into a single parameter that can explain a 
variety of experiments [Fig.~\ref{fig:fig_4}(a)]. Nonlinear dependence of $JMR$ ratios on the impurity concentrations are shown [Fig.~\ref{fig:fig_4}(b-d)] 
in difference with earlier linear estimates from empirical models\cite{Moodera1}. Accordingly, small differences in spin-states/concentrations of magnetic 
impurities are shown to cause large deviations in $JMRs$. Interesting similarities and differences among devices having Pd, Ni and Co impurities are 
pointed out, which could be signatures of the spin states of oxidized Pd and Ni impurities and low-spin/high-spin phase transitions for oxidized cobalt 
impurities.

\begin{acknowledgments}
This work was supported by the MARCO focus center for Materials, Structure and Devices and the NSF Network for Computational Nanotechnology. Part of this work was carried 
out at the Jet Propulsion Laboratory under a contract with the National Aeronautics and Space Administration NASA funded by ARDA.
\end{acknowledgments}
\text{~~~}

\appendix*

\section{\label{app:main} Spin Dephasing Self Energy} 

In the following, the scattering tensors stated in Eqs.~(\ref{eq:spin_flip}-\ref{eq:no_spin_scat}) of the main paper are obtained starting from 
standard expressions obtained in the self-consistent Born approximation from the NEGF formalism. Here we start from the formulation in Ref.~\onlinecite{Datta} 
(see Sections 10.4 and A.4) which represents a generalization of the earlier treatments \cite{Gerhard,Datta_Condensed_Matter,Mahan,DattaFermi}. 
For spatially localized scatterers we have:

\begin{subequations}
\label{eq:scat_tensor_definition}
\begin{eqnarray}
D_{\sigma _i \sigma _j ;\sigma _k \sigma _l }^n \left( {t,t'} \right) = \left\langle {\sum\limits_{s_\beta  ,s_\alpha  } {\tau _{\sigma _i s_\alpha  ;\sigma _k s_\beta  } \left( t \right)\tau _{\sigma _j s_\alpha  ;\sigma _l s_\beta  }^* \left( {t'} \right)} } \right\rangle  \nonumber \\ 
= \left\langle {\sum\limits_{s_\alpha  ,s_\beta  } {\left\langle {s_\alpha  } \right|H_{I;\sigma _i \sigma _k } \left( t \right)\left| {s_\beta  } \right\rangle \left\langle {s_\beta  } \right|H_{I;\sigma _l \sigma _j }^\dag  \left( {t'} \right)\left| {s_\alpha  } \right\rangle } } \right\rangle  
\text{~~~~~}\label{eq:Dn_tau}
\end{eqnarray}
\begin{eqnarray}
D_{\sigma _i \sigma _j ;\sigma _k \sigma _l }^p \left( {t,t'} \right) = \left\langle {\sum\limits_{s_\beta  ,s_\alpha  } {\tau _{\sigma _l s_\beta  ;\sigma _j s_\alpha  } \left( t \right)\tau _{\sigma _k s_\beta  ;\sigma _i s_\alpha  }^* \left( {t'} \right)} } \right\rangle  \nonumber \\ 
= \left\langle {\sum\limits_{s_\alpha  ,s_\beta  } {\left\langle {s_\beta  } \right|H_{I;\sigma _l \sigma _j } \left( t \right)\left| {s_\alpha  } \right\rangle \left\langle {s_\alpha  } \right|H_{I;\sigma _i \sigma _k }^\dag  \left( {t'} \right)\left| {s_\beta  } \right\rangle } } \right\rangle  
\text{~~~~~}\label{eq:Dp_tau} 
\end{eqnarray}
\end{subequations}
where  $\left| s_\alpha  \right\rangle$ and $\left| s_\beta  \right\rangle$ are \textit{impurity spin subspace} states and $H_{I}$ 
is the interaction Hamiltonian [Eq.~(\ref{eq:hamiltonian})] defined within the channel \textit{electron spin subspace} as:

\begin{subequations}
\label{eq:HI_definition}
\begin{eqnarray} 
H_{I;\sigma _i \sigma _k } \left( {t} \right) = \left\langle {\sigma _i } \right|H_I \left( {t} \right)\left| {\sigma _k } \right\rangle 
\end{eqnarray}
\begin{eqnarray}
H_{I;\sigma _l \sigma _j } \left( {t} \right) = \left\langle {\sigma _l } \right|H_I \left( {t} \right)\left| {\sigma _j } \right\rangle 
\end{eqnarray}
\begin{eqnarray}
H_{I;\sigma _i \sigma _k }^\dag  \left( {t'} \right) = \left\langle {\sigma _i } \right|H_I^\dag  \left( {t'} \right)\left| {\sigma _k } \right\rangle 
\end{eqnarray}
\begin{eqnarray}
H_{I;\sigma _l \sigma _j }^\dag  \left( {t'} \right) = \left\langle {\sigma _l } \right|H_I^\dag  \left( {t'} \right)\left| {\sigma _j } \right\rangle 
\end{eqnarray}
\end{subequations}
For an uncorrelated impurity spin ensemble with:
\begin{equation} 
\rho  = \sum\limits_{s_\alpha  } {w _{s_\alpha  } \left| {s_\alpha  } \right\rangle \left\langle {s_\alpha  } \right|}  = \sum\limits_{s_\beta  } {w _{s_\beta  } \left| {s_\beta  } \right\rangle \left\langle {s_\beta  } \right|} 
\end{equation}
averaging in Eqs.~(\ref{eq:scat_tensor_definition}a-\ref{eq:scat_tensor_definition}b) can be done through a weighted summation of spin scattering 
rates of magnetic impurities:

\begin{widetext}
\begin{subequations}
\label{eq:scat_tensor_definition_operator_form}
\begin{eqnarray} 
D_{\sigma _i \sigma _j ;\sigma _k \sigma _l }^n \left( {t,t'} \right) &=& \sum\limits_{s_\alpha  ,s_\beta  } {w _{s_\beta  } \left\langle {s_\alpha  } \right|H_{I;\sigma _i \sigma _k } \left( t \right)\left| {s_\beta  } \right\rangle {\rm  }\left\langle {s_\beta  } \right|H_{I;\sigma _l \sigma _j }^\dag  \left( {t'} \right)\left| {s_\alpha  } \right\rangle } \nonumber \\ 
&=& \sum\limits_{s_\alpha  } {\left\langle {s_\alpha  } \right|H_{I;\sigma _i \sigma _k } \left( t \right){\rm  } [\rho] H_{I;\sigma _l \sigma _j }^\dag  \left( {t'} \right)\left| {s_\alpha  } \right\rangle }  
= tr\left( {\rho {\rm  }H_{I;\sigma _l \sigma _j }^\dag  \left( {t'} \right){\rm  }H_{I;\sigma _i \sigma _k } \left( t \right)} \right) 
\end{eqnarray}
\begin{eqnarray}
D_{\sigma _i \sigma _j ;\sigma _k \sigma _l }^p \left( {t,t'} \right) &=& \sum\limits_{s_\alpha  ,s_\beta  } {w _{s_\alpha  } \left\langle {s_\beta  } \right|H_{I;\sigma _l \sigma _j } \left( t \right)\left| {s_\alpha  } \right\rangle {\rm   }\left\langle {s_\alpha  } \right|H_{I;\sigma _i \sigma _k }^\dag  \left( {t'} \right)\left| {s_\beta  } \right\rangle } \nonumber  \\ 
&=& \sum\limits_{s_\beta  } {\left\langle {s_\beta  } \right|H_{I;\sigma _l \sigma _j } \left( t \right){\rm  } [\rho] {\rm  }H_{I;\sigma _i \sigma _k }^\dag  \left( {t'} \right)\left| {s_\beta  } \right\rangle }  
= tr\left( {\rho {\rm  }H_{I;\sigma _i \sigma _k }^\dag  \left( {t'} \right){\rm  }H_{I;\sigma _l \sigma _j } \left( t \right)} \right) 
\end{eqnarray}
\end{subequations}
\end{widetext} 
The trace $tr\left( {\rho A} \right)$ for any operator $A$ is independent of representation. Accordingly, $[D^n]/[D^p]$ 
scattering tensors can be evaluated using any convenient basis for magnetic impurity spin states.

Through Jordan-Wigner transformation, single spins can be thought as an empty or singly occupied fermion state:

\begin{subequations}
\label{eq:jordan_wigner}
\begin{equation}
\left|  \uparrow  \right\rangle  \equiv a^ +  \left| 0 \right\rangle 
,\label{subeq:creation}
\end{equation}
\begin{equation}
\left|  \downarrow  \right\rangle  \equiv \left| 0 \right\rangle 
,\label{subeq:annihilation}
\end{equation}
\end{subequations}
with \textit{creation/annihilation} operators for the channel electrons:

\begin{subequations}
\label{eq:jordan_wigner_operators}
\begin{equation}
a^ + (t)  = \sigma ^ + (t)   = \left[ {\begin{array}{*{20}c}
   0 & {e^{i\omega_e t} }  \\
   0 & 0  \\
\end{array}} \right]
,\label{subeq:creation_operator}
\end{equation}
\begin{equation}
a(t) = \sigma ^ -  (t) = \left[ {\begin{array}{*{20}c}
   0 & 0  \\
   {e^{ - i\omega_e t} } & 0  \\
\end{array}} \right]
.\label{subeq:annihilation_operator}
\end{equation}
\end{subequations}
For degenerate electron spin states ($\hbar \omega _e  = 0$), there's no time dependence  as such ${a^\dag  \left( t \right) \to a^\dag  }$  , $ {a\left( t \right) \to a}$.

Pauli spin matrices are related to creation/annihilation operators through $\sigma_x=(a^\dag+a)/2$, $\sigma_y=(a^\dag+a)/2i$ and  $\sigma_z=a^\dag a- 1/2$. 
Accordingly, interaction Hamiltonian $H_{I}\left( {\vec r,t} \right) = J{\rm  }\delta \left( {\vec r - \vec R} \right){\rm  }\vec \sigma  \cdot \vec S\left( t \right)$ 
can be expressed as:

\begin{widetext}
\begin{equation}
H_{{\mathop{\rm I}} } \left( {\vec r,t} \right) = J\delta \left( {\vec r - \vec R} \right)\left[ {\frac{1}{2}aS_ +  \left( t \right) + \frac{1}{2}a^ \dag  S_ -  \left( t \right) + \left( {a^ \dag  a - \frac{1}{2}} \right)S_Z \left( t \right)} \right]
\\,\label{eq:hamiltonian_sec}
\end{equation}
\end{widetext}
where S is the spin operator for the localized magnetic impurity. 

Substituting the interaction Hamiltonian from Eq.~(\ref{eq:hamiltonian_sec}) into Eqs.~(\ref{eq:scat_tensor_definition_operator_form}a-\ref{eq:scat_tensor_definition_operator_form}b) will yield:
\begin{widetext}
\begin{subequations}
\label{eq:scat_tensor_expanded}
\begin{equation}
\begin{array}{*{20}c}
\begin{array}{l}
 \left| {\sigma _k \sigma _l } \right\rangle  \to  \\ 
 \text{~~~~~~~~~~} \left\langle {\sigma _i \sigma _j } \right| \downarrow  \\ 
 \end{array} & {\begin{array}{*{20}c}
   {\left| { \uparrow  \uparrow } \right\rangle }              \text{~~~~~~~~~~~~~~}
& {\left| { \downarrow  \downarrow } \right\rangle }           \text{~~~~~~~~~~~~~~}
& {\left| { \uparrow  \downarrow } \right\rangle }             \text{~~~~~~~~~~~~~~~}
& {\left| { \downarrow  \uparrow } \right\rangle }  \\         \text{~~~~~~~~~~~}
\end{array}}  \\
   {H_{I;\sigma _l \sigma _j }^\dag \left( {t'} \right)H_{I;\sigma _i \sigma _k } \left( t \right){\rm   } 
= {\rm } \text{~~} \begin{array}{*{20}c}
   {\left\langle { \uparrow  \uparrow } \right| }  \\
   {\left\langle { \downarrow  \downarrow } \right| }  \\
   {\left\langle { \uparrow  \downarrow } \right| }  \\
   {\left\langle { \downarrow  \uparrow } \right| }  \\
\end{array}} & {\left[ {\begin{array}{*{20}r}
   { {S_Z \left( t' \right)S_Z \left( {t} \right)} }   & { {S_ +  \left( t' \right)S_ -  \left( {t} \right)} }   & {   {S_ +  \left( t' \right)S_Z \left( {t} \right)} }   & {   {S_Z \left( t' \right)S_ -  \left( {t} \right)} }  \\
   { {S_{\rm -}  \left( t' \right)S_{\rm +}  \left( {t} \right)} } & { {S_Z \left( t' \right)S_Z \left( {t} \right)} } & { - {S_Z \left( t' \right)S_ +  \left( {t} \right)} }   & { - {S_ -  \left( t' \right)S_Z \left( {t} \right)} }  \\
   { {S_-  \left( t' \right)S_Z \left( {t} \right)} }  & {-{S_Z \left( t' \right)S_ -  \left( {t} \right)} }     & { - {S_Z \left( t' \right)S_Z \left( {t} \right)} }     & {   {S_ -  \left( t' \right)S_ -  \left( {t} \right)} }  \\
   { {S_Z \left( t' \right)S_+  \left( {t} \right)} }  & {-{S_ +  \left( t' \right)S_Z \left( {t} \right)} }     & {   {S_ +  \left( t' \right)S_ +  \left( {t} \right)} } & { - {S_Z \left( t' \right)S_Z \left( {t} \right)} }  \\

\end{array}} \right]}  \\
\end{array}
,\label{eq:tensor_n_operator}
\end{equation}
\begin{equation}
\begin{array}{*{20}c}
   \begin{array}{l}
 \left| {\sigma _k \sigma _l } \right\rangle  \to  \\ 
 \text{~~~~~~~~~~} \left\langle {\sigma _i \sigma _j } \right| \downarrow  \\ 
 \end{array} & {\begin{array}{*{20}c}
   {\left| { \uparrow  \uparrow } \right\rangle }              \text{~~~~~~~~~~~~~~}
& {\left| { \downarrow  \downarrow } \right\rangle }           \text{~~~~~~~~~~~~~~}
& {\left| { \uparrow  \downarrow } \right\rangle }             \text{~~~~~~~~~~~~~~~}
& {\left| { \downarrow  \uparrow } \right\rangle }  \\         \text{~~~~~~~~~~~}
\end{array}}  \\
   {H_{I;\sigma _i \sigma _k }^\dag \left( {t'} \right)H_{I;\sigma _l \sigma _j } \left( t \right){\rm   } = {\rm  } \text{~~} \begin{array}{*{20}c}
   {\left\langle { \uparrow  \uparrow } \right| }  \\
   {\left\langle { \downarrow  \downarrow } \right| }  \\
   {\left\langle { \uparrow  \downarrow } \right| }  \\
   {\left\langle { \downarrow  \uparrow } \right| }  \\
\end{array}} & {\left[ {\begin{array}{*{20}r}
   { {S_Z \left( t' \right)S_Z \left( {t} \right)} }     & {  {S_ -  \left( t' \right)S_ +  \left( {t} \right)}  } & {   {S_Z \left( t' \right)S_ +  \left( {t} \right)} }    & {   {S_ -  \left( t' \right)S_Z \left( {t} \right)} }  \\
   { {S_ +  \left( t' \right)S_ -  \left( {t} \right)} } & {  {S_Z \left( t' \right)S_Z \left( {t} \right)}} & { - {S_ +  \left( t' \right)S_Z \left( {t} \right)} }    & { - {S_Z \left( t' \right)S_ -  \left( {t} \right)} }  \\
   { {S_Z \left( t' \right)S_ -  \left( {t} \right)} }   & { -{S_ -  \left( t' \right)S_Z \left( {t} \right)} }    & { - {S_Z \left( t' \right)S_Z \left( {t} \right)}   }    & {   {S_ -  \left( t' \right)S_ -  \left( {t} \right)} }  \\
   { {S_ +  \left( t' \right)S_Z \left( {t} \right)} }   & { -{S_Z \left( t' \right)S_ +  \left( {t} \right)} }    & {   {S_ +  \left( t' \right)S_ +  \left( {t} \right)} }  & { - {S_Z \left( t' \right)S_Z \left( {t} \right)} }  \\
\end{array}} \right]}  \\
\end{array}
,\label{eq:tensor_p_operator}
\end{equation}
\end{subequations}

\end{widetext}

The localized magnetic impurity spin-operators can be written in its diagonalized \textit{impurity spin-subspace} as:
\begin{subequations}
\label{eq:jordan_wigner_operators}
\begin{equation}
S_ +   = d^ \dag   = \left[ {\begin{array}{*{20}c}
   0 & {e^{i\omega _q t} }  \\
   0 & 0  \\
\end{array}} \right]
\end{equation}
\begin{equation}
S_ -   = d = \left[ {\begin{array}{*{20}c}
   0 & 0  \\
   {e^{ - i\omega _q t} } & 0  \\
\end{array}} \right]
\end{equation}
\begin{equation}
S_Z  = d^ \dag  d - \frac{1}{2} = \frac{1}{2}\left[ {\begin{array}{*{20}r}
   1 & 0  \\
   0 & { - 1}  \\
\end{array}} \right]
\end{equation}
\end{subequations}
with $\omega _q  = {{\Delta E_I } \mathord{\left/ {\vphantom {{\Delta E_I(\omega_q) } \hbar }} \right. \kern-\nulldelimiterspace} \hbar }$ 
where $\Delta E_I$ is the energy difference between spin-up and spin-down states for the localized magnetic impurities. 

For a given impurity density matrix of the form (Fu+Fd=1):  

\begin{equation}
\rho  = N_I(\omega_q) \left[ {\begin{array}{*{20}c}
   {F_u } & 0   \\
   {0 } & {F_d }  \\
\end{array}} \right]
,\label{eq:rho_matrix}
\end{equation}
with $N_I$ being total number of impurities, the desired quantities $ {{\left[ {D^n } \right]} \mathord{\left/ {\vphantom {{\left[ {D^n } \right]} {\left[ {D^p } \right]}}} \right. \kern-\nulldelimiterspace} {\left[ {D^p } \right]}}$ 
can be obtained by evaluating the expectation values of the operators in Eqs.~(\ref{eq:tensor_n_operator}) and (\ref{eq:tensor_p_operator}). 
Here the only non-zero elements are:

\begin{subequations}
\label{eq:expect_val}
\begin{equation}
 tr\left( {\rho S_Z \left( {t'} \right)S_Z \left( t \right)} \right) = \frac{1}{4} 
,\label{eq:expect_1}
\end{equation}
\begin{equation}
 tr\left( {\rho S_ +  \left( {t'} \right)S_ -  \left( t \right)} \right) = F_u e^{ - i\omega _q \left( {t - t'} \right)}  
,\label{eq:expect_2}
\end{equation}
\begin{equation}
 tr\left( {\rho S_ -  \left( {t'} \right)S_ +  \left( t \right)} \right) = F_d e^{i\omega _q \left( {t - t'} \right)}  
.\label{eq:expect_3}
\end{equation}
\end{subequations}

Finally, for a given impurity density matrix [Eq~(\ref{eq:rho_matrix})], $[D^n]/[D^p]$ tensors are obtained as:
\begin{widetext}

\begin{subequations}
\begin{equation}
\begin{array}{*{20}c}
   {\begin{array}{l}
 \left| {\sigma _k \sigma _l } \right\rangle  \to  \\ 
 \text{~~~~~} \left\langle {\sigma _i \sigma _j } \right| \downarrow  \\ 
 \end{array} } & {\begin{array}{*{20}c}
\text{~~~~~}
   {\left| { \uparrow  \uparrow } \right\rangle }                 \text{~~~~~~~~~~}
& {\left| { \downarrow  \downarrow } \right\rangle }              \text{~~~~~~~~}
& {\left| { \uparrow  \downarrow } \right\rangle }                \text{~~~~}
& {\left| { \downarrow  \uparrow } \right\rangle }  \\            \text{~~~~~}
\end{array}}  \\
   {{\rm D}^{\rm n} \left( {t,t'} \right) = \sum\limits_{\omega_q} {J^2 N{}_I(\omega_q)} {\rm  } \text{~~} \begin{array}{*{20}c}
    {\left\langle { \uparrow  \uparrow } \right| }  \\
    {\left\langle { \downarrow  \downarrow } \right| }  \\
    {\left\langle { \uparrow  \downarrow } \right| }  \\
    {\left\langle { \downarrow  \uparrow } \right| }  \\
\end{array}} & {\left[ {\begin{array}{*{20}c}
   {{1 \mathord{\left/ {\vphantom {1 4}} \right. \kern-\nulldelimiterspace} 4}} & {F_u e^{-i\omega_q \left( {t - t'} \right)} } & 0 & \text{~~}0  \\
   {F_d e^{i\omega_q \left( {t - t'} \right)} } & {{1 \mathord{\left/ {\vphantom {1 4}} \right. \kern-\nulldelimiterspace} 4}} & 0 & \text{~~} 0  \\
   0 & 0 & { - {1 \mathord{\left/ {\vphantom {1 4}} \right. \kern-\nulldelimiterspace} 4}} & \text{~~} 0  \\
   0 & 0 & 0 & \text{~~} { - {1 \mathord{\left/ {\vphantom {1 4}} \right. \kern-\nulldelimiterspace} 4}}  \\
\end{array}} \right]}  \\
\end{array}
,\label{eq:time_domain_Dn}
\end{equation}

\begin{equation}
\begin{array}{*{20}c}
   {\begin{array}{l}
 \left| {\sigma _k \sigma _l } \right\rangle  \to  \\ 
 \text{~~~~~} \left\langle {\sigma _i \sigma _j } \right| \downarrow  \\ 
 \end{array} } & {\begin{array}{*{20}c}
\text{~~~~~}
   {\left| { \uparrow  \uparrow } \right\rangle }                 \text{~~~~~~~~~~}
& {\left| { \downarrow  \downarrow } \right\rangle }              \text{~~~~~~~~}
& {\left| { \uparrow  \downarrow } \right\rangle }                \text{~~~~}
& {\left| { \downarrow  \uparrow } \right\rangle }  \\            \text{~~~~~}
\end{array}}  \\
   {{\rm D}^{\rm p} \left( {t,t'} \right) = \sum\limits_{\omega_q} {J^2 N{}_I(\omega_q)} {\rm  } \text{~~} \begin{array}{*{20}c}
    {\left\langle { \uparrow  \uparrow } \right| }  \\
    {\left\langle { \downarrow  \downarrow } \right| }  \\
    {\left\langle { \uparrow  \downarrow } \right| }  \\
    {\left\langle { \downarrow  \uparrow } \right| }  \\
\end{array}} & {\left[ {\begin{array}{*{20}c}
   {{1 \mathord{\left/ {\vphantom {1 4}} \right. \kern-\nulldelimiterspace} 4}} & {F_d e^{i\omega_q \left( {t - t'} \right)} } & 0 &  \text{~~} 0  \\
   {F_u e^{-i\omega_q \left( {t - t'} \right)} } & {{1 \mathord{\left/ {\vphantom {1 4}} \right.  \kern-\nulldelimiterspace} 4}} &  0 & \text{~~} 0  \\
   0 & 0 & { - {1 \mathord{\left/ {\vphantom {1 4}} \right. \kern-\nulldelimiterspace} 4}} & \text{~~} 0  \\
   0 & 0 & 0 & \text{~~} { - {1 \mathord{\left/  {\vphantom {1 4}} \right. \kern-\nulldelimiterspace} 4}}  \\
\end{array}} \right]}  \\
\end{array}
,\label{eq:time_domain_Dp}
\end{equation}
\end{subequations}
\end{widetext}

It's convenient to work with the Fourier transformed functions as such $(t - t') \to \hbar \omega $:
\begin{equation}
e^{ - i\omega _q \left( {t - t'} \right)} e^{{{ - \eta \left| {t - t'} \right|} \mathord{\left/
 {\vphantom {{ - \eta \left| {t - t'} \right|} \hbar }} \right.
 \kern-\nulldelimiterspace} \hbar }}  \to \delta \left( {\hbar \omega  - \hbar \omega _q } \right)
\end{equation}
where $\eta$ being a positive infinitesimal. With Fourier transforming Eqs.~(\ref{eq:time_domain_Dn}-\ref{eq:time_domain_Dp}) will simplify to Eqs.~(\ref{eq:spin_flip}-\ref{eq:no_spin_scat}) \cite{DattaFermi}. 
For the calculations reported in this article, diagonal elements not leading to spin-dephasing are omitted due to their negligible effect on $JMR$ ratios. 
In this case $[D^n]/[D^p]$ scattering tensors simplifies to a form (Eq.~(\ref{eq:tensor_relation})) which can be understood from simple common-sense 
arguments (Eqs.~(\ref{eq:ioscat_up}) and (\ref{eq:ioscat_down})).

\bibliography{yanik_dephasing}

\end{document}